\documentclass[runningheads]{llncs}
\usepackage[T1]{fontenc}
\usepackage{graphicx}
\usepackage{hyperref}
\usepackage{color}

\usepackage{cite}
\usepackage{amsmath,amssymb,amsfonts}
\usepackage{algorithmic}
\usepackage{textcomp}
\usepackage{xcolor}

\begin{document}

\bibliographystyle{splncs04}
\title{Dynamic Data-Driven Digital Twin Testbed for Enhanced First Responder Training and Communication}
\titlerunning{Dynamic Data-Driven Digital Twin Testbed}
%
\author{Alyssa Cassity\orcidID{0000-1111-2222-3333} \and
Hieu Le\orcidID{0000-0003-2510-073X} \and
Hernan Santos\orcidID{0009-0005-0113-5067} \and
Erik Priest\orcidID{2222--3333-4444-5555} \and
Jian Tao\orcidID{0000-0003-1374-8678}}
\authorrunning{A. Cassity et al.}
%
\institute{Texas A\&M University, College Station, Texas, USA 
\\
\email{\{aac0428,hieult,hsantos,epriest,jtao\}@tamu.edu}
}
\maketitle              
\begin{abstract}
The study focuses on developing a digital twin testbed tailored for public safety technologies, incorporating simulated wireless communication within the digital world. The integration enables rapid analysis of signal strength, facilitating effective communication among personnel during catastrophic incidents in the virtual environment. The virtual world also helps with the training of first responders. The digital environment is constructed using the actual training facility for first responders as a blueprint. Using the photo-reference method, we meticulously constructed all buildings and objects within this environment. These reconstructed models are precisely placed relative to their real-world counterparts. Subsequently, all structures and objects are integrated into the Unreal Engine (UE) to create an interactive environment tailored specifically to the requirements of first responders.

\end{abstract}
\section{Introduction}
\label{sec:intro}
The public safety sector recognizes the urgent need for first responders to have access to advanced communication and networking technologies, similar to those commonly used by consumers on commercial networks. However, implementing these new technologies into public safety operations involves more than just acquiring equipment and services.

Firstly, it is crucial to provide comprehensive training for first responders and emergency managers. This training is essential to maximize the benefits of new devices and technologies, while also managing the added stress of handling additional equipment. Additionally, in certain situations, equipment may need to be redesigned or customized to effectively address various emergency scenarios.

A potential solution to these challenges is the creation of a testbed that integrates Virtual Reality (VR) systems with Digital Twin (DT) and IoT technologies. This virtual testbed not only facilitates the training of public safety professionals, but also allows for controlled testing of equipment and interfaces. In doing so, it reduces the risk of first responders and training costs. Meanwhile, it simplifies the testing of equipment for operational suitability. In essence, the virtual testbed complements traditional field tests and serves as a valuable tool to improve public safety operations.

Secondly, effective communication between first responders is crucial for public safety operations, as it determines their ability to collaborate and complete tasks efficiently. In extreme cases, it can be a matter of life or death. Therefore, integrating wireless communication features, especially coverage maps, into a digital twin (DT) environment is essential.

Significant progress has been made in wireless communication simulation, particularly in enhancing communication quality and data transmission rates. A key contribution is the open source library Sionna, which uses parallel computing to simulate digital communication processes \cite{sionna}. Sionna's differentiable capability enables the integration of deep learning into digital communication, thereby improving performance. Researchers have utilized Sionna to explore synchronization in IoT networks \cite{aoudia2022deep} and deep learning-based channel coding \cite{cammerer2022graph, gong2023graph}. Additionally, Sionna's ray-tracing methods support the analysis of environmental properties and building structures \cite{hoydis2023sionna}. 

In the study by \cite{alkhateeb2023real}, wireless communication simulation is incorporated into a DT for wireless networks. This research envisions real-time updated DTs using multi-modal sensing data from distributed infrastructure and user devices to guide communication and sensing decisions. Enabled by advancements in 3D mapping, multi-modal sensing, ray tracing, and machine learning, the study presents methods for creating and utilizing these real-time digital twins, explores their applications and challenges, and presents a research platform for exploring diverse research directions related to digital twins. Unlike this study, our current work integrates radio coverage map generation into our DT system, which provides visual representations of signal strength throughout the environment.

Recently, the Dynamic Data Driven Applications Systems (DDDAS) paradigm has demonstrated substantial influence across a range of critical application domains and systems operating under dynamic conditions. This paradigm supports "systems-analytics" and "Dynamic Digital Twin" capabilities. In alignment with the DDDAS approach, our DT system integrates instrumentation data into models in real-time, enabling dynamic control and acquisition of the data.

In this paper, we present our work to integrate DT into public safety operations in a physical training facility named Disaster City \cite{teex}. Based on our previous work \cite{nicole2023}, our objective is to develop a testbed enhanced with digital twin technologies, incorporating state-of-the-art user interface/user experience innovations and advanced wireless simulation models. This initiative is designed to provide the public safety community with a photorealistic virtual reality environment, allowing them to evaluate and engage with the latest sensing and communication technologies.


\section{Methodology}
\label{sec:methodology}

Disaster City spans 52 acres and is strategically positioned adjacent to the TEEX Brayton Fire Training Field \cite{teex}. This extensive training center accommodates the diverse skill set needs of modern emergency response professionals.

The development of digital twins for first responders involves several phases. Firstly, a virtual representation of the physical environment is constructed on the basis of real-world observations and measurements of buildings and objects. This virtual model is then utilized to generate radio coverage maps, allowing for the estimation of wireless signal strength. Given the critical importance of signal strength for first responders, this information helps to determine effective communication between team members.

In addition, sensors are deployed throughout the actual facility and connected to a server to continuously monitor environmental conditions, such as temperature and weather. All data gathered from these sensors, along with the generated coverage map and building reconstruction, are seamlessly integrated into Unreal Engine (UE) 5, which is a powerful graphics engine, to create interactive environments for users.

\subsection{Framework}

\begin{figure*}[tbp]
  \centering
  \includegraphics[width=1\linewidth]{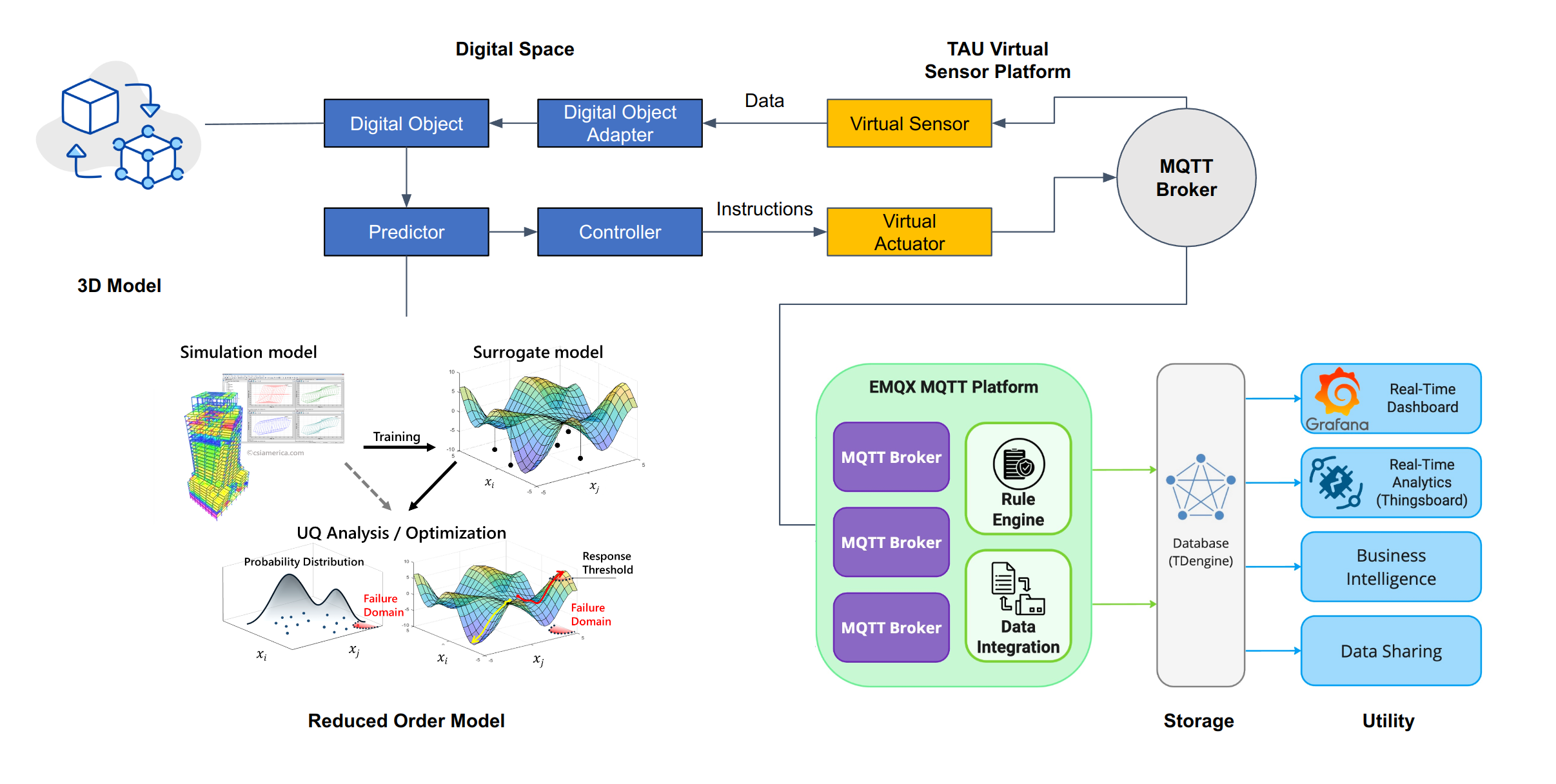}
  \caption{Digital Twin framework for Disaster City}
  \label{figure:dt_framework}
\end{figure*}

The general framework of the project is shown in Figure \ref{figure:dt_framework}. Data collection from the physical environment is facilitated through an extensive network of sensors and Internet of Things (IoT) devices. These streams of data are then stored in a database and processed by an MQTT (Message Queuing Telemetry Transport) Broker, a system specifically designed to efficiently transmit IoT data to a centralized computer system \cite{mqtt}. Subsequently, the processed information is channeled through a virtual sensor platform to prepare data for integration into the Digital Space.

In this digital realm, sophisticated data analytics leverages highly precise 3D geometric models, facilitating the processing of all incoming data. To enhance the efficacy of the digital twin system, advanced machine learning algorithms and reduced order models can be integrated into the digital space to improve predictive modeling and decision-making capabilities. The predictive results are then transmitted to a virtual controller, which can issue commands to physical entities via virtual actuators and the MQTT broker. 

This integrated framework, illustrated in Figure \ref{figure:dt_framework}, facilitates seamless interaction between the physical and digital domains, enabling real-time analysis and control of the Disaster City infrastructure. The entire digital twin system functions as a DDDAS model, incorporating real-time feedback loops between the physical and digital environments to enhance responsiveness and adaptability to changing conditions. The dynamic feedback loop automatically updates all properties of the virtual environment, with the results from the virtual environment being transmitted back to the physical world to issue operational commands. Consequently, the virtual environment continuously mirrors the physical world, while the physical instruments can be optimally controlled by predictions and commands generated based on the virtual environment.


\subsection{Interactive Environment in Unreal Engine}

\begin{figure}[tbp]
  \centering
  \includegraphics[width=1\linewidth]{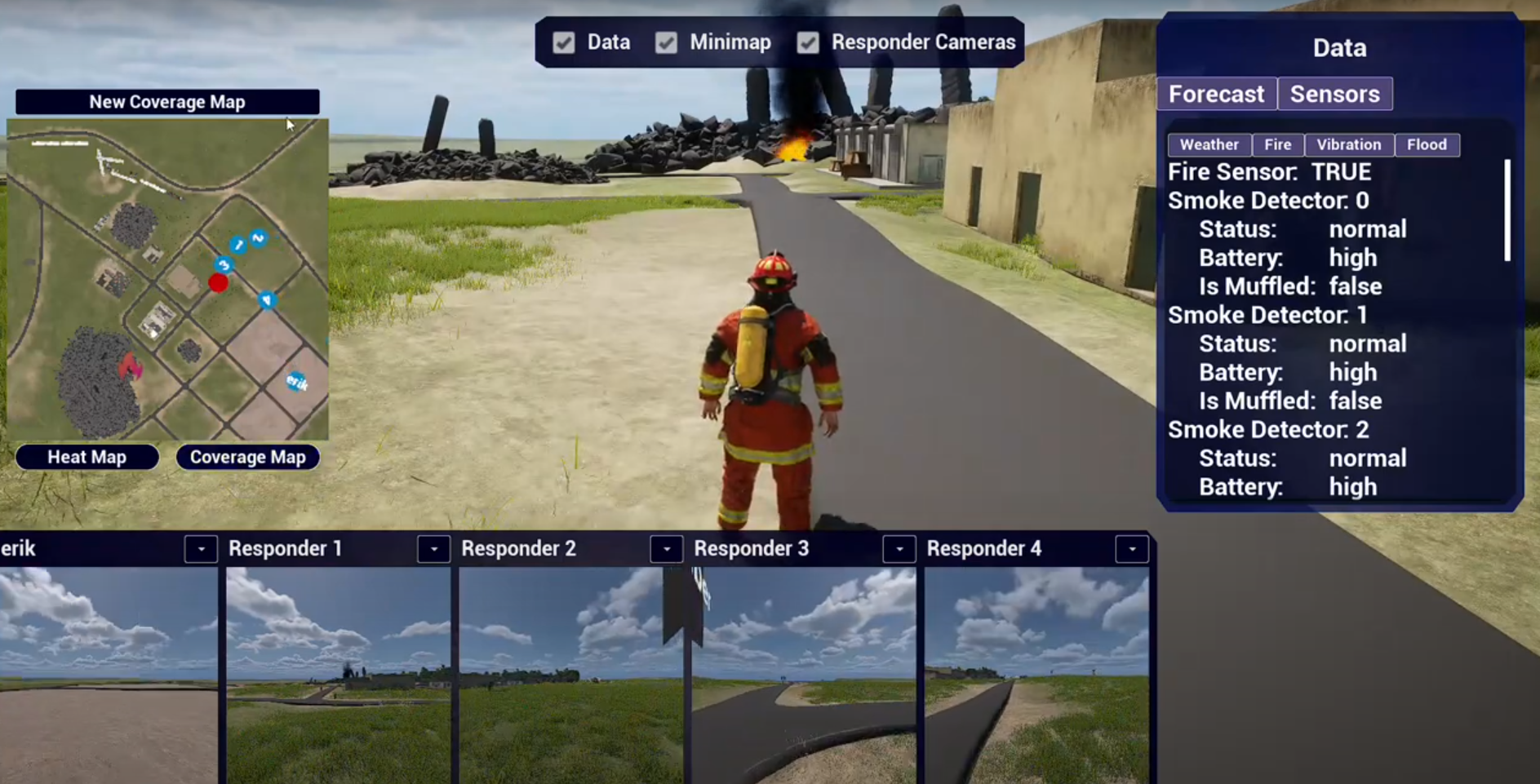}
  \caption{Disaster City virtual environment with different types of weather information displayed on the user interface.}
  \label{figure:disaster_city_environment}
\end{figure}
Given the expansive size of Disaster City and its various training fields, each simulating different catastrophic events, modeling the entire complex in a virtual environment is both labor-intensive and challenging. Accurate geometric measurements of all structures within the facility are essential and are obtained through on-site assessments. This step is crucial to ensure fidelity and precision in the virtual representation of Disaster City's infrastructure. Following field observations, we use photo references to meticulously construct each building and assets for the Disastery City (see Fig. \ref{figure:disaster_city_environment}).



\subsection{Integration with Disaster City IoT sensors}

Originally created for UE 4, an MQTT plugin has been modified to seamlessly operate within UE 5. This plugin streamlines connectivity to the MQTT server and offers a range of functionalities, such as topic subscription and transmission of messages containing different data formats. This framework facilitates the transmission of various topics across the server, guaranteeing real-time data processing within the UE environment.

Upon receiving a message, an event is activated within UE, leading to the conversion of the message into a string format. Subsequent development progresses according to the specific requirements of the task at hand. Understanding the structure and content of the received message is crucial, as accurate data processing is essential to parse the message into the appropriate variables tailored for the virtual environment. These variables form the foundation for scene modification, facilitating dynamic adjustments such as alterations to weather conditions and lighting within the scene, thereby enhancing its realism and interactivity. 

\subsection{Radio Coverage Maps using Sionna}

We integrate wireless radio signals into the virtual environment constructed using UE. This objective is achieved through a sequential process that involves two key tasks: first, the creation of coverage maps using Sionna \cite{sionna} based on a predefined configuration file provided through an MQTT broker, followed by the integration of these generated coverage maps into the UE environment via the MQTT broker. This approach ensures the seamless integration of wireless communication elements into the virtual world, enhancing its realism and utility for wireless communication applications.

The first step in creating coverage maps involves performing external procedures outside of the UE environment. This includes exporting the virtual environment to a specialized file format necessary for Sionna. Prior to export, the scene configuration is established to accurately determine the coordinates of the Access Point (AP). Following this setup, both the configuration file and the exported scene are transmitted to a simulation-capable system via the MQTT broker. Afterwards, the exported model is simulated within Sionna to generate radio coverage maps according to the pre-determined configuration.

Leveraging advanced ray tracing capabilities of Sionna specifically designed for wireless communication scenarios, we produce coverage maps that illustrate path gains for a specific 3D model. The ray tracing method emulates real-world scenarios in which a transmission source emits rays that interact with surrounding objects, bounce off them indefinitely until their intensity diminishes. 

Internally, Sionna utilizes Monte Carlo approximation techniques to calculate the received signal power across a specified area. At an elevation of 1.5 meters above ground level, a 2D grid is created, consisting of numerous small units called cells. Each cell serves as a container for gathering received power values at its center. When a ray intersects a cell within the grid, the received power is computed and added to the cumulative value within that cell. Mathematically, path gain for coverage maps generated from Sionna are computed based on equation (\ref{eq:integral_cmap}). The integral over the cell $C_{i,j}$ can be approximated by Monte Carlo approximation \cite{sionna}.

\begin{equation}
b_{i,j} = \frac{1}{|C|} \int_{C_{i,j}}{|h(s)|^2ds},
\label{eq:integral_cmap}
\end{equation}
\\
where $b_{i,j}$ is the path gain while $C_{i,j}$ represents any given cell containing intersecting rays within itself, with $i$ and $j$ denoting its coordinates within a 2D grid; $|h(s)|^2$ is the square of the amplitude of path coefficients at a position $s=(x,y)$ within the cell $C_{i,j}$; $ds$ stands for the infinitesimally small surface element, expressed as $ds = dx \cdot dy$.

This process of computation and accumulation continues until either the maximum specified number of bounces is reached or the ray fails to intersect any objects in the scene. After the computation terminates, the received power values within each cell are combined to generate a single value representing the total received power at the center of that cell.

Following the creation of coverage maps by Sionna, these maps are converted into a suitable format for transmission via the MQTT broker to UE. Upon receiving the radio coverage maps at UE, the data undergoes processing through UE render targets. These render targets act as 2D coordinate planes within UE's environment and have the capability for dynamic drawing during runtime. The data delivered via MQTT provides detailed information about the location and intensity of each point on a heatmap. Subsequently, the coordinate system from the coverage map is converted to 2D coordinates.

To initiate the generation of a new coverage map for a different location, the previous render target displaying the heatmap is presented on the screen. Users have the capability to engage with the map by clicking anywhere to set a new location of the radio tower. The coordinates of the new point selected by the user's click are subsequently converted back into the coordinates of the virtual world map. These coordinates are then transmitted via MQTT to initiate the generation of a new map
. This systematic procedure enables real-time visualization and interaction within the simulated space.



\section{Results and Discussion}
\label{sec:results_and_discussion}

\subsection{Coverage Maps}

\begin{figure}[tbp]
  \centering
  \includegraphics[width=1\linewidth]{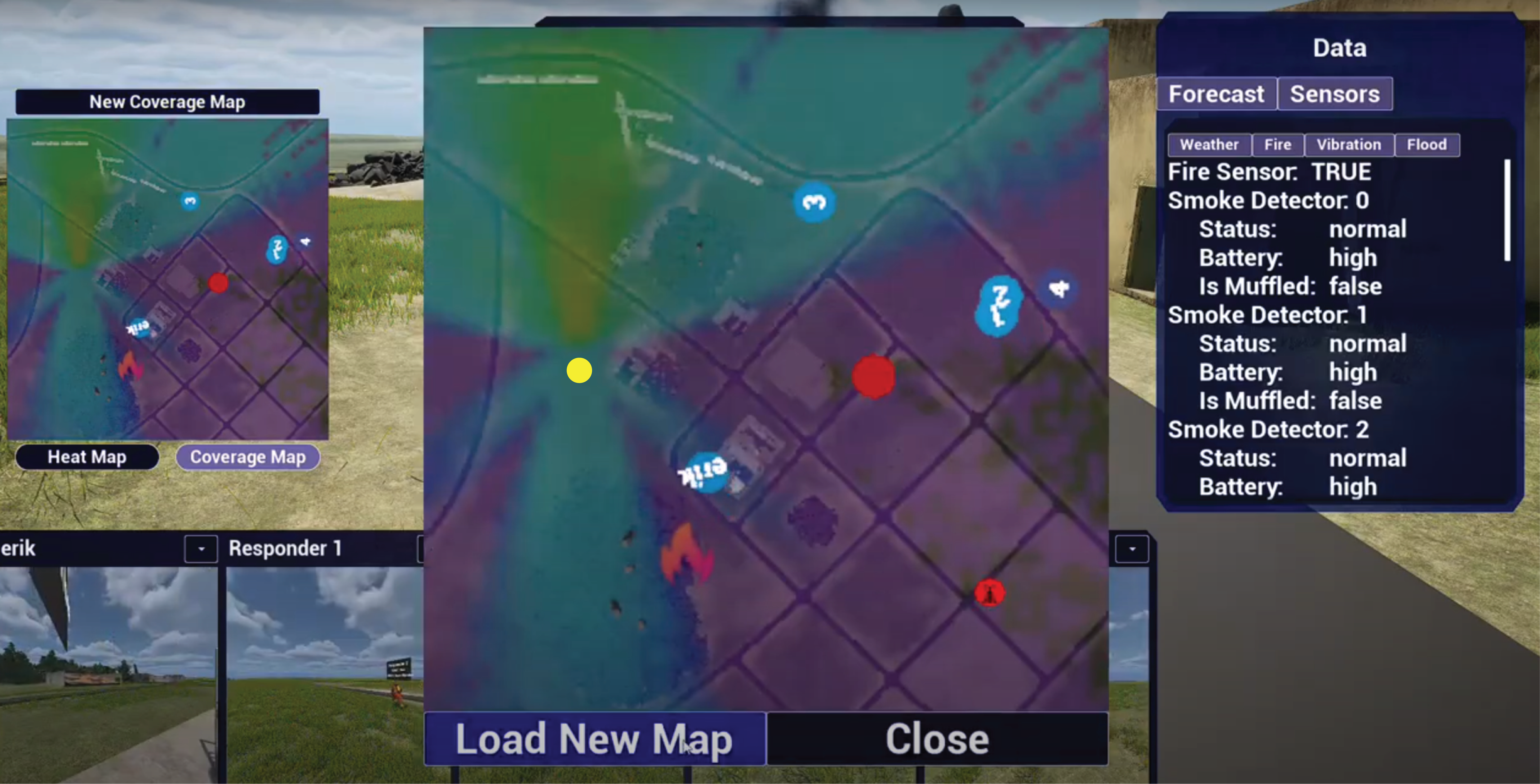}
  \caption{Coverage map for the specified radio tower location, indicated by a yellow circle, in Disaster City.}
  \label{figure:disaster_city_cmap}
\end{figure}

Figure \ref{figure:disaster_city_cmap} shows the coverage map generated based on the position of the radio tower, marked by a yellow circle on the left side of the map. This map illustrates the path gain throughout the area, with yellowish regions indicating high path gain and purple regions indicating low path gain. The figure shows a higher path gain in one direction, which is due to the transmitter being equipped with a directional antenna oriented northward.

The location and orientation of the radio tower can be adjusted. When the radio tower configuration changes, the coverage map updates accordingly. Generating a new coverage map takes approximately 9 seconds. As this simulation runs in the background, users can continue interacting with the environment without interruption. Given that changes to the radio tower configuration are infrequent, this simulation time is considered reasonable.

\section{Conclusion}
\label{sec:conclusion}

In this study, we developed a digital twin testbed tailored for public safety technologies, integrating simulated wireless communication to enhance signal strength analysis and first responder training. By reconstructing an actual first responder training facility in the Unreal Engine, we created a highly realistic and interactive environment. This virtual world not only allows for the optimization of communication strategies in disaster scenarios but also significantly improves the preparedness of first responders through immersive training exercises. The testbed's design, based on precise photo-reference methods, ensures a close replication of real-world conditions, thereby enhancing the effectiveness of training and operational strategies. Although our DT has not yet undergone a thorough validation process, we intend to address this by incorporating actual physical activities from first responders in our follow-up work.

This advancement represents a leap forward in public safety technology, offering a scalable and adaptable platform for future innovations. Ultimately, this digital twin testbed stands as a critical tool in improving emergency response capabilities, potentially saving lives by ensuring first responders are well-prepared and efficiently connected during catastrophic incidents.

\begin{credits}
\subsubsection{\ackname} The authors would like to thank Walt Magnussen, Nick Duffield, Andre Thomas, Narasimha Reddy, Michael Fox, Derek Ladd, Patrick Newman, and Renold A.J. from Texas A\&M University, Jason Moats and Ray Ivie from Texas A\&M Engineering Extension Service for their suggestions and comments. This work was partially supported by the awards 70NANB22H004 and 70NANB22H074 from U.S. Department of Commerce, National Institute of Standards and Technology. Tao is thankful for the support from the Texas A\&M Institute of Data Science through the Career Initialization Fellow Program and the Data Science Course Development Grant Program.

\subsubsection{\discintname}
The authors have no competing interests to declare that are relevant to the content of this article.

\subsubsection{Data Availability}
We are working on a website \href{https://dtkit.org/}{dtkit.org} as part of the effort to release the code. The 3D assets cannot be released, but the codebase will be made publicly accessible.

\end{credits}
%
%
%

\bibliography{./main_ref}

\begin{thebibliography}{1}
\providecommand{\url}[1]{\texttt{#1}}
\providecommand{\urlprefix}{URL }
\providecommand{\doi}[1]{https://doi.org/#1}

\bibitem{alkhateeb2023real}
Alkhateeb, A., Jiang, S., Charan, G.: Real-time digital twins: Vision and research directions for 6g and beyond. IEEE Communications Magazine  (2023)

\bibitem{aoudia2022deep}
Aoudia, F.A., Hoydis, J., Cammerer, S., Van~Keirsbilck, M., Keller, A.: Deep learning-based synchronization for uplink nb-iot. In: GLOBECOM 2022-2022 IEEE Global Communications Conference. pp. 1478--1483. IEEE (2022)

\bibitem{cammerer2022graph}
Cammerer, S., Hoydis, J., Aoudia, F.A., Keller, A.: Graph neural networks for channel decoding. In: 2022 IEEE Globecom Workshops (GC Wkshps). pp. 486--491. IEEE (2022)

\bibitem{gong2023graph}
Gong, A., Cammerer, S., Renes, J.M.: Graph neural networkrs for enhanced decoding of quantum ldpc codes. arXiv preprint arXiv:2310.17758  (2023)

\bibitem{nicole2023}
Hatch, N., Magnussen, W., Tao, J.: Efforts towards a digital twin-based testbed for public safety. In: Proceedings of Cyber-Physical Systems and Internet of Things Week 2023. p. 297–299. CPS-IoT Week '23, Association for Computing Machinery, New York, NY, USA (2023). \doi{10.1145/3576914.3588017}, \url{https://doi.org/10.1145/3576914.3588017}

\bibitem{hoydis2023sionna}
Hoydis, J., Aoudia, F.A., Cammerer, S., Nimier-David, M., Binder, N., Marcus, G., Keller, A.: Sionna rt: Differentiable ray tracing for radio propagation modeling. arXiv preprint arXiv:2303.11103  (2023)

\bibitem{sionna}
Hoydis, J., Cammerer, S., {Ait Aoudia}, F., Vem, A., Binder, N., Marcus, G., Keller, A.: {Sionna: An Open-Source Library for Next-Generation Physical Layer Research}. arXiv preprint  (Mar 2022)

\bibitem{mqtt}
MQTT: Mq telemetry transport (2024), \url{https://mqtt.org/}

\bibitem{teex}
TEEX: Texas a\&m engineering extension service (2024), \url{https://teex.org/}

\end{thebibliography}

\end{document}